%Paper: hep-th/9211086
%From: SEIF@ITSICTP.BITNET
%Date: Thu, 19 Nov 92 11:27 N

%%%%%%%%%%%%%%%%%%%%%%%%%%%%%%%%%%%%%%%%%%%%%%%%%%%%%%%%%%%%%%%%%%%%%%%%%
%    THIS IS ICTP.TEX - 1992 Version -
%    ictp.tex is a file that contains definitions created for typing
%    ictp preprints and internal reports when using TEX.
%    Definitions can be added, but PLEASE DO NOT make any variations
%    to the existing definitions!
%%%%%%%%%%%%%%%%%%%%%%%%%%%%%%%%%%%%%%%%%%%%%%%%%%%%%%%%%%%%%%%%%%%%%%%%%
%
\magnification=1200
\vsize=24truecm
 \nopagenumbers
 \footline={ \ifnum\pageno = 1 \primapag\else\altrepag\fi}
      \def\primapag{\hss\ \hss}
       \def\altrepag{\hss\folio\hss}
\def\IC#1{\null\vskip-1.5truecm
{\rightline{IC/92/#1}\bigskip\bigskip
\centerline{International Atomic Energy Agency}
\smallskip\centerline{and}\smallskip
\centerline{United Nations Educational Scientific
and Cultural Organization}\medskip\centerline{INTERNATIONAL CENTRE FOR
THEORETICAL
PHYSICS}\vskip 2truecm  }}

\def\TITLE#1{{\centerline{\bf #1}}
}

\def\MIRAMARE{\par\vfill {\centerline{MIRAMARE -- TRIESTE}}\medskip
             \centerline{\today}\vfill \eject}

\def\AUTHOR#1{{\centerline {#1}}\smallskip}

\def\ABSTRACT{{\centerline{ABSTRACT}}\bigskip}

\parindent=40pt

  \baselineskip=14pt
  \parskip=7pt plus 1pt

\def\today{\ifcase\month\or
             January\or February\or March \or April\or May\or June\or
             July\or August\or September\or October\or November\or December\fi
             \ \number\year}

%**************************************************************
%               tagged sec numbers
% **************************************************************
\global\newcount\secno \global\secno=0
\global\newcount\meqno \global\meqno=1
\global\newcount\subsecno \global\subsecno=0
\def\SECTION#1{\global\advance\secno by1\global\meqno=1\global\subsecno=0
\bigbreak   \bigskip    % (combination \goodbreak\bigskip\bigskip)
\noindent \hangindent 40pt {\bf\hbox to 40pt{\the\secno.\hfil}
           #1}\par\nobreak
                 \medskip\nobreak\message{#1 , }}
\def\SUBSECTION#1{\global\advance\subsecno by1\medbreak
      \noindent\hangindent 40pt
     {\bf\hbox to 40pt{\the\secno.\the\subsecno\hfil}#1}\par\nobreak
                 \medskip\nobreak\message{#1 ,  }}

% **************************************************************
%       \eqn\label{a+b=c}       gives a displayed equation with number
%                               chosen consecutively within sections.
%     \eqnn and \eqna define labels in advance
% **************************************************************
%\newwrite\efile \let\firsteqn=T
%\def\writeqno#1%
% {\if T\firsteqn \immediate\openout\efile=eqns.tmp\global\let\firsteqn=F\fi%
%\immediate\write\efile{#1 \string#1}\global\advance\meqno by1}

%\def\eqnn#1{\xdef #1{(\the\secno.\the\meqno)}\writeqno#1}
%\def\eqna#1{\xdef #1##1{(\the\secno.\the\meqno##1)}\writeqno{#1{}}}

%\def\eqn#1#2{\xdef #1{(\the\secno.\the\meqno)}%
%$$#2\eqno(\the\secno.\the\meqno)$$\writeqno#1}

% **************************************************************
%                        footnotes
% **************************************************************
\global\newcount\ftno \global\ftno=1
\def\FOOT#1{\footnote{$^{\the\ftno}$}{#1}\ %
\global\advance\ftno by1}

% **************************************************************
%     \fig\label{text}
% generates a number, assigns it to \label, generates an entry.
% To list the figs on a separate page,  \listfigs
% **************************************************************
\global\newcount\figno \global\figno=1
\newwrite\ffile
\def\fig#1#2{\the\figno\nfig#1{#2}}
\def\nfig#1#2{\xdef#1{\the\figno}%
\ifnum\figno=1\immediate\openout\ffile=figs.tmp\fi%
\immediate\write\ffile {\noexpand \item{Fig. \noexpand#1 :\ }\noexpand#2}%
\global\advance\figno by1}
\def\semi{;\hfil\noexpand\break}

% **************************************************************
%     \table\label{text}
% generates a number, assigns it to \label, generates an entry.
% To list the tables on a separate page,  \listtable
% **************************************************************
\global\newcount\tableno \global\tableno=1
\newwrite\ffile
\def\table#1#2{\the\tableno\ntable#1{#2}}
\def\ntable#1#2{\xdef#1{\the\tableno}%
\ifnum\tableno=1\immediate\openout\ffile=table.tmp\fi%
\immediate\write\ffile {\noexpand \item{Table. \noexpand#1 :\ }\noexpand#2}%
\global\advance\tableno by1}

% **************************************************************
%     \ref\label{text}
% generates a number, assigns it to \label, generates an entry.
% To list the refs on a separate page,  \listrefs
% **************************************************************
\global\newcount\refno \global\refno=1
\newwrite\rfile
\def\ref#1#2{$^{[\the\refno]}$\nref#1{#2}}
\def\nref#1#2{\xdef#1{$^{[\the\refno]}$}%
\ifnum\refno=1\immediate\openout\rfile=refs.tmp\fi%
\immediate\write\rfile{\noexpand\item{\noexpand#1\ }\noexpand#2.}%
\global\advance\refno by1}

% **************************************************************
% this is for the captions
% **************************************************************

% *************************************************************
% This is for tagged  references
% ************************************************************+

% **************************************************************
% this is to make tables
% **************************************************************

% **************************************************************
% and finally acknowledgments
% **************************************************************

% **************************************************************
% And REFERENCES
% **************************************************************

% **************************************************************
% Some useful abbreviations
% **************************************************************
\def\mc{\,\raise -2.truept\hbox{\rlap{\hbox{$\sim$}}\raise5.truept
\hbox{$<$}\ }}
\def\Mc{\,\raise -2.truept\hbox{\rlap{\hbox{$\sim$}}\raise5.truept
\hbox{$>$}\ }}
%%%%%%%%%%%%%%%%%%%%%%%%%%%%%%%%%%%%%%%%%%%%%%%%%%%%%%%%%%%%%%%%%%%%%%%%
%                     DEFINITIONS ADDED
%%%%%%%%%%%%%%%%%%%%%%%%%%%%%%%%%%%%%%%%%%%%%%%%%%%%%%%%%%%%%%%%%%%%%%%%
%
%  for acknowledgments with one author or more
%

%
%  for section headings, subsection headings, theorems, lemmas, proofs, etc.
%

%
% to center the place and date on the front page
%
\def\MIRAMARE#1{\vfill\centerline{MIRAMARE -- TRIESTE}\medskip
         \centerline{#1}\vfill}
%
%  a small square
%
\def\square{\sqcap\kern-6pt\lower2.4pt\hbox{--}\ }
\def\sq{\sqcap\kern-8pt\lower2.4pt\hbox{--}\ }
%
% The following are used to type ``blackboard bold'' letters
%
\def\bbone{{\rm 1}\kern-4pt{\rm 1}}

\def\bbc{{\rm C}\kern-6pt\hbox{\vrule height6.5pt width0.8pt}\ \, }

\def\bbg{{\rm G}\kern-5pt\hbox{\vrule height6pt width 0.8pt}\ \, }

\def\bbo{{\rm O}\kern-4.8pt\hbox{\vrule height6.5pt width0.8pt}\  }

\def\bbq{{\rm Q}\kern-5pt\hbox{\vrule height6pt width 0.7pt}\  \, }

\def\bbs{{\rm S}\kern-3.5pt\hbox{\vrule height6.5pt width 0.7pt}\ }
\def\bbz{{\rm Z}\!\!\! {\rm Z}}
\def\subbbc{{\rm C}\kern-3.5pt\hbox{\vrule height4.5pt width0.4pt}\, }
% The next definition is for reference numbers typed superscript with
% a bracket after the number
%

%

%                   For Footnotes
%

%
%
%                   New definition \frac (for fractions)
%
\def\frac#1/#2{\leavevmode\kern.1em
\raise.5ex\hbox{\the\scriptfont0 #1}
\kern-.1em/\kern-.15em\lower.25ex\hbox{\the\scriptfont0 #2}}
%

%						Definition for a black box

%
%    since \hbar does not work with MacTex
%    use  \barh for normal sized text
%         \ssbarh for superscripts
%          \ovssbarh for superscripts with fractions
\def\barh{h\kern-5pt\raise3pt\hbox{-}\ }
\def\ssbarh{h\kern-4.5pt\raise3pt\hbox{-}\,}
\def\ovssbarh{h\kern-3.5pt\hbox{-}\,}

%%%%%%%%%%%%%%%%%%%%
%    FOR TABLES
%%%%%%%%%%%%%%%%%%

\def\longrightharpoonup{-\kern-3pt\hbox{$\rightharpoonup$}\ }

\def\centerpar{
\let\endgraf=\par \edef\restorehsize{\hsize=14truecm}
\def\par{\endgraf \restorehsize \let\par=\endgraf}
\advance\hsize by-\parindent
\item{}}

\hfuzz=15pt
\def\ux{\underline x}
\def\highgamma{\raise 2pt\hbox{$\gamma$}}
\def\ba{\bar\alpha}
\def\bb{\bar\beta}
\def\bs{\bar\sigma}
\def\bg{\bar\gamma}
\def\bd{\bar\delta}
\def\a{\alpha}
\def\b{\beta}
\def\g{\gamma}
\def\s{\sigma}
\def\d{\delta}
\def\l{\lambda}
\def\v{\varepsilon}
\def\bp{p\kern-6.5pt\lower 2.5pt\hbox{$-$}}
\def\ha{\hat\alpha}
\def\hb{\hat\beta}
\def\hg{\hat\gamma}

\IC{354}
\TITLE{THE RENORMALIZATION GROUP FOR FLAG MANIFOLDS}
\vskip1.5truecm
\AUTHOR{S. Randjbar--Daemi\quad and\quad J. Strathdee}
\centerline{International Centre for Theoretical Physics, Trieste, Italy.}
\vskip1truecm
\ABSTRACT

The renormalization group equations for a class of non--relativistic
quantum $\sigma$--models targeted on flag manifolds are given. These
models emerge in a continuum limit of generalized Heisenberg
antiferromagnets. The case of the ${SU(3)\over U(1)\times U(1)}$
manifold is studied in greater detail. We show that at zero temperature
there is a fixed point of the RG transformations in
$(2+\varepsilon )$--dimensions where the theory becomes relativistic.
We study the linearized RG transformations in the vicinity of this
fixed point and show that half of the couplings are irrelevant. We also
show that at this fixed point there is an enlargement of the global
isometries of the target manifold. We construct a discrete
non--abelian enlargement of this kind.

\MIRAMARE{October 1992}
\vfill\eject

\SECTION{INTRODUCTION}

In an earlier paper $^{1)}$ the notion of a generalized spin system was
examined. These systems are defined by associating the generators
of a Lie algebra, $G$, in some finite dimensional matrix representation,
with the sites of a lattice. The Hamiltonian is expressed as a sum of terms
comprising various $G$--invariant couplings between generators on
different sites. The idea was to study models which go beyond the well
known $SU(2)$ spin systems $^{2)}$ to see if the enriched group structure
has interesting consequences. For example, the topological
term discovered by Haldane $^{3)}$ in the continuum approximation to the
Heisenberg chain, that suggests a qualitative distinction between the
integer-- and half--integer spin models: how does it generalize? Or the
statements concerning the existence or not of an ordered ground state,
depending on the spin $^{2)}$: do they generalize? In the previous paper,
apart from kinematical matters, only the large quantum number,
correspondence theory, limit was considered together with the naive long
wavelength behaviour. A generalization to the case of $S(N)$ models of
the Holstein--Primakoff formulae was obtained but the main result was
a general description of the continuum limit in the classical
approximation. Depending on assumptions about the nature of the ground
state configuration (ferromagnetic, antiferromagnetic, or some more
general type of order) it was found that the resulting classical field
theory could be formulated as a generalized $\sigma$--model. In these
models the field variables take their values on a coset manifold
$G/H$ which generally turns out to be a so--called flag manifold
(where $H$ is the maximal Abelian subgroup of $G$). These models are
$G$--invariant and non--relativistic. Our intention now is to examine
the simplest of these flag manifold $\sigma$--models with a view to
finding fixed points of the renormalization group equations.

We start with a $\sigma$--model Lagrangian of the general form
$${\cal L} ={1\over 2}\ g_{\mu\nu}(\phi )\ \partial_t\phi^\mu
  \partial_t\phi^\nu +{1\over 2}\ k_{\mu\nu}(\phi )\
  \partial_i\phi^\mu\partial_i\phi^\nu
\eqno(1.1)$$
where the fields $\phi^\mu (t,\ux )$ are targeted on some coset space,
$G/H$. The base space is flat $D+1$--dimensional Euclidean spacetime.
For simplicity we assume $O(D)$ isotropy in space, but not relativistic
invariance. The tensors $g_{\mu\nu}$ and $k_{\mu\nu}$ are supposed to
be the most general $G$--invariant, positive definite tensors that can
be assigned to the target manifolds. Later we shall specialize to the case
of flag manifolds.

The coefficient tensors in (1.1) can be specified in terms of a finite
number of parameters. For example, in a frame basis, $e_\mu\ ^\alpha
(\phi )$, they take the form
$$\eqalignno{
  g_{\mu\nu}(\phi ) &=e_\mu\ ^\alpha (\phi )\ e_\nu\ ^\beta (\phi )\
  g_{\alpha\beta}\cr
  k_{\mu\nu}(\phi ) &=e_\mu\ ^\alpha (\phi )\ e_\nu\ ^\beta (\phi )\
  k_{\alpha\beta}              &(1.2)\cr}$$
where the coefficients $g_{\alpha\beta}$ and $k_{\alpha\beta}$ are
$\phi$--independent and $H$--invariant. These tensors comprise the
coupling parameters of the model and it is their evolution under the
action of the renormalization group that we wish to study.

In the literature there is an extensive study of two--dimensional
relativistic $\sigma$--models $^{4)}$ especially in connection with
string theory $^{5)}$. Our model is non--relativistic and we are primarily
interested in the quantum statistics of this model which can be regarded
as a generalization of the work of Chakravarty {\it et al.} $^{6)}$
from the case of $G/H={SU(2)\over U(1)}$ to more general $SU(N)$
flag manifolds.

In more than one space dimension the Lagrangian (1.1) is not ultraviolet
renormalizable. This is not a serious consideration, however, since the
model has its origin in a lattice system which is necessarily ultraviolet
finite. The ultraviolet pathologies of (1.1) will therefore be suppressed
by imposing a cutoff at the order of the lattice spacing. This is not an
entirely trivial matter, however, since one is changing the topology
of momentum space. This space is toroidal (and compact) in the lattice
case whereas it becomes a ball in the cutoff continuum case. To
maintain $G$--invariance in such cases it is usually more appropriate
to employ some $G$--invariant regularization procedure. We shall avoid
most of the difficulties by using a covariant background--field method
$^{7)}$ for computing quantum corrections.

To define the action of the renormalization group we use the momentum
shell technique $^{8)}$. This means integrating out the hard components,
those with momenta in the shell
$$\Lambda\geq\vert\ \bp\vert >{\Lambda\over s}\qquad (s>1)$$
where $\Lambda$ is the ultraviolet cutoff. The resulting effective
Lagrangian for the soft components must retain the $G$--invariant
form (1.1) after an appropriate field redefinition. But the coupling
parameters will be modified,
$$\eqalign{
  g_{\alpha\beta}\to g_{\alpha\beta} &+\Delta g_{\alpha\beta}\cr
  k_{\alpha\beta}\to k_{\alpha\beta} &+\Delta k_{\alpha\beta}\ .\cr}$$
The aim is to obtain the effective $g_{\alpha\beta}$ and $k_{\alpha\beta}$
as functions of $s$ and study their behaviour in the limit $s\to\infty$.

In Sec.2 we describe the background field method as it applies to our
problem and compute the 1--loop contributions to the effective
couplings. These can be expressed quite generally in terms of the
structure constants of the invariance group, $G$. The general form of the
evolution equations, and their restriction to the case of flag manifolds
is obtained. In these computations we impose periodicity, $t\to t+\beta$, in
the Euclidean time so that temperature becomes one of the variables.
(It will be assumed that the temperature is low so that expansions in the
time derivative $\partial_0\phi$ are meaningful.)

As a consequence of the compact topology of the $t$--direction a new
gauge invariant term is induced in the effective action which disappears
in the zero temperature limit. This term is non--local in the time
variable and it is constructed from the pull--back of the holonomy of
the spin--connection of $G/H$ (i.e. the Wilson line) to the spacetime
manifold.

The results of Sec.2 are further specialized to the $SU(N)$--flag
manifolds in Sec.3. Here the $\beta$--functions assume a simple form
and it becomes possible to show that, in the limit of zero temperature
there exists a non--trivial fixed point at which the theory becomes
``relativistic'' in the sense that $k_{\mu\nu} =c^2g_{\mu\nu}$ and all
modes propagate with the same velocity, $c$. We examine the behaviour
of trajectories in the neighbourhood of this fixed point and determine
the critical indices (i.e. the eigenvalues of the matrix which governs the
linearized RG equations) associated with it. The result is that half of
the couplings are relevant. This shows that the system would maintain a
relativistic form under renormalization, which is not surprising. But
it also shows, disappointingly, that the relativistic fixed point would
not be reached if the initial structure were not already relativistic.
At the critical point there is an enlargement of the isometry group of
the target space to include discrete transformations. This is
illustrated for the case of $SU(N)$ flag manifolds in Sec.4 where it is
shown that the permutation group, $S_N$, is incorporated. However, the
action of $S_N$ is not free, there are fixed points. The discussion of
some technical matters have been relegated to the two appendices at
the end of the paper.

\SECTION{THE BACKGROUND FIELD METHOD}

To take advantage of the $G$--invariance of the system it is advisable
to set up a manifestly covariant scheme for computations. One such is the
so--called covariant background field method $^{7)}$. This method involves
the introduction of a set of geodesic normal coordinates on the target
space. It is useful in perturbative calculations. The original field
variables $\phi^\mu (x)$ are replaced by expansions in powers of new
fields $\xi^\mu (x)$ -- the normal coordinates -- which are treated as
small quantities
$$\phi^\mu (x)=\bar\phi^\mu (x) +\xi^\mu (x) +\dots   \eqno(2.1)$$
where $\bar\phi^\mu (x)$ is the background field and is treated as an
external field. The terms of the series (2.1) are determined by solving
a geodesic type of equation,
$${\partial^2\highgamma^\mu\over\partial u^2} +
  {\partial\highgamma^\lambda\over\partial u}\
  {\partial\highgamma^\nu\over\partial u}\
  \Gamma_{\nu\lambda}\ ^\mu (\highgamma )=0                 \eqno(2.2)$$
where $\highgamma^\mu (x,u)$ interpolates between $\bar\phi^\mu (x)$
and $\phi^\mu (x)$, i.e.
$$\eqalignno{
  \highgamma^\mu (x,0) &=\bar\phi^\mu (x)\cr
  \highgamma^\mu (x,1) &=\phi^\mu (x)\ .   &(2.3)\cr}$$
The form of the expansion (2.1) depends on the choice of connection,
$\Gamma_{\nu\lambda}\ ^\mu$, used in the geodesic equation (2.2).
One possible choice for $\Gamma$ would be the Riemannian connection
corresponding to some metric on $G/H$. However, our Lagrangian (1.1)
involves two independent ``metric'' tensors, $g_{\mu\nu}$ and $k_{\mu\nu}$,
and it is quite unclear which to choose. The best connection, in the sense of
being the most convenient, is the one with respect to which both
$g_{\mu\nu}$ and $k_{\mu\nu}$ are covariantly constant. It is uniquely
defined.

When (2.1) is substituted into (1.1) one obtains the expansion,
$$\eqalignno{
  {\cal L} &= {1\over 2}\ g_{\mu\nu}\ \partial_t\bar\phi^\mu\
  \partial_t\bar\phi^\nu\cr
  &\quad +g_{\mu\nu}\ \partial_t\bar\phi^\mu (\nabla_t\xi^\nu +
  \xi^\lambda\ \partial_t\bar\phi^\rho\ T_{\lambda\rho}\ ^\nu )\cr
  &\quad +{1\over 2}\ g_{\mu\nu}\ \nabla_t\xi^\mu\ \nabla_t\xi^\nu +
  \left( T_{\nu\rho}\ ^\lambda\ g_{\lambda\mu} +{1\over 2}\
  T_{\nu\mu}\ ^\lambda\ g_{\lambda\rho}\right)\xi^\nu\
  \nabla_t\xi^\mu\ \partial_t\bar\phi^\rho\cr
  &\quad +{1\over 2} \left( -R_{\rho\mu\nu}\ ^\tau\ g_{\tau\sigma} +
  g_{\rho\tau}\ T_{\nu\sigma}\ ^\lambda\ T_{\mu\lambda}\ ^\tau +
  g_{\lambda\tau}\ T_{\nu\rho}\ ^\lambda\ T_{\mu\sigma}\ ^\tau
  \right)\xi^\mu\xi^\nu\partial_t\bar\phi^\rho\partial_t
  \bar\phi^\sigma\cr
  &\quad +O(\xi^3)\cr
  &\quad + ({\rm terms\ with}\ \ \partial_t\to\partial_j,\ \
  g_{\mu\nu}\to k_{\mu\nu})             &(2.4)\cr}$$
where all tensors, $g_{\mu\nu},R_{\rho\mu\nu}\ ^\tau$, etc. are
evaluated on the background, $\bar\phi$. Details of the derivation of
this formula are given in Appendix A.

The vector $\xi^\mu$ can be referred to the background frame basis,
$e_\mu\ ^\alpha (\bar\phi )$,
$$\xi^\mu =\xi^\alpha\ e_\alpha\ ^\mu\ .          \eqno(2.5)$$
In this basis the covariant derivatives are given by
$$\nabla_t\ \xi^\alpha =\partial_t\xi^\alpha +
  \xi^\beta\ \omega_{t\beta}\ ^\alpha\ ,              \eqno(2.6)$$
etc., where the spin connection, $\omega (\bar\phi )$, is defined in
Appendix A. In the frame basis the curvature and torsion tensors
associated with this connection are particularly simple. They are
$\bar\phi$--independent and they are expressed in terms of structure
constants of $G$,
$$\eqalignno{
  R_{\alpha\beta\gamma}\ ^\delta &= c_{\alpha\beta}\ ^{\bar\sigma}\
  c_{\gamma\bar\sigma}\ ^\delta\cr
  T_{\alpha\beta}\ ^\gamma &=-c_{\alpha\beta}\ ^\gamma\ . &(2.7)\cr}$$
In these formulae the labels $\alpha ,\beta\dots $ refer to the tangent
space of $G/H$ while $\bar\sigma$ refers to the algebra of $H$. In
terms of generators,
$$\eqalignno{
  [Q_{\ba}, Q_{\bb}] &= c_{\ba\bb}\ ^{\bg}\ Q_{\bg}\cr
  [Q_\alpha ,Q_{\bb}] &= c_{\alpha\bb}\ ^\gamma\ Q_\gamma\cr
  [Q_\alpha ,Q_\beta ] &=c_{\alpha\beta}\ ^{\bg}\
  Q_{\bg} +c_{\alpha\beta}\ ^\gamma\ Q_\gamma\ .   &(2.8)\cr}$$

To compute the 1--loop contribution to $\Delta g_{\alpha\beta}$ and
$\Delta k_{\alpha\beta}$ the bilinear terms in the expansion (2.4) are
sufficient. These terms determine the propagator, $<\xi^\alpha\xi^\beta >$,
and the couplings to the external field, $\bar\phi$. One needs to
evaluate the graphs of Fig.1. These are vacuum graphs with
respect to the quantum fields $\xi^\alpha$, but the
momentum integrations are\break
\vskip4truecm

\item{Fig.1}
The 1--loop contributions to the effective Lagrangian. Solid lines
represent the propagator $<\xi^\alpha\xi^\beta >$. Dashed lines
represent the external field, $\partial\bar\phi^\mu\ e_\mu\ ^\alpha
(\bar\phi )$. The external field is slowly varying and these graphs
represent the terms of second order in the small quantities,
$\partial\bar\phi^\mu$.

\noindent  resticted to the shell,
$$\Lambda\geq \vert \bp\vert >\Lambda /s    \ .               \eqno(2.9)$$
Since one wants only the long wavelength, low frequency part of the
effective action, it is enough to pick out only the second order terms
in $\partial\bar\phi^\mu$ to identify the coefficients $\Delta g_{\mu\nu}$
and $\Delta k_{\mu\nu}$, i.e.
$${\cal L}_{eff} ={\cal L} +{1\over 2}\ \Delta g_{\mu\nu}\
  \partial_t\bar\phi^\mu\ \partial_t\bar\phi^\nu + {1\over 2}\
  \Delta k_{\mu\nu}\ \partial_j\bar\phi^\mu\ \partial_j
  \bar\phi^\nu +\dots
\eqno(2.10)$$
where the dots indicate non--local and higher order terms. Among the
non--local terms there will be gauge invariant terms constructed from the
pull--back of the holonomy of the spin--connection of the target
manifold. Such terms will contribute only at non--zero $T$ when the
time direction describes a circle of radius $\beta ={1\over T}$. They
will disappear in the limit $\beta\to\infty$.

For the contributions of graphs (a) and (b) one obtains the expressions
$$\eqalignno{
  \Delta_a\ g_{\alpha\beta} &=\left( -R_{\alpha\gamma\delta}\
  ^\varepsilon\ g_{\varepsilon\beta} +g_{\alpha\varepsilon}\
  T_{\delta\beta}\ ^\varepsilon\ T_{\gamma\sigma}\ ^\varepsilon +
  g_{\varepsilon\sigma}\ T_{\delta\alpha}\ ^\varepsilon\
  T_{\gamma\beta}\ ^\sigma\right) G^{\gamma\delta}\cr
  \Delta_a\ k_{\alpha\beta} &=\left( -R_{\alpha\gamma\delta}\
  ^\varepsilon\ k_{\varepsilon\beta} +k_{\alpha\varepsilon}\
  T_{\delta\beta}\ ^\varepsilon\ T_{\gamma\sigma}\ ^\varepsilon +
  k_{\varepsilon\sigma}\ T_{\delta\alpha}\ ^\varepsilon\
  T_{\gamma\beta}\ ^\sigma\right) G^{\gamma\delta}\cr
  \Delta_b\ g_{\a\b} &=-\left( T_{\a\g}\ ^\v\ g_{\v\d}-{1\over 2}\
  T_{\g\d}\ ^\v\ g_{\v\a}\right)\left( T_{\b\g '}\ ^{\v '}\
  g_{\v '\d '} -{1\over 2}\ T_{\g '\d '}\ ^{\v '}\
  g_{\v '\b}\right) K^{\g\d\g '\d '}_{tt}\cr
  \d_{ij}\ \Delta_b\ k_{\a\b} &=-\left( T_{\a\g}\ ^\v\
  k_{\v\d} -{1\over 2}\ T_{\g\d}\ ^\v\ k_{\v\a}\right)
  \left( T_{\b\g '}\ ^{\v '}\ k_{\v '\d '}-{1\over 2}\
  T_{\g' \d '}\ ^{\v '}\  k_{\v '\b}\right)
  K^{\g\d\g '\d '}_{ij}   \cr
  &                   &(2.11)\cr}$$
where
$$\eqalignno{
  G^{\g\d} &= {1\over\b}\ \sum_r\int\left( {d\bp\over 2\pi}\right)^D\
  G^{\g\d}(p)\cr
  K^{\g\d\g '\d '}_{\mu\nu} &={1\over\b}\ \sum_r\int\left(
  {d\bp\over 2\pi}\right)^D\ p_\mu p_\nu\left( G^{\g\g '}(p)\
  G^{\d\d '}(-p)-G^{\g\d '}(p)\ G^{\d\g '}(-p)\right)\ .   &(2.12)\cr}$$
The propagator, $G^{\alpha\beta}(p)$, is defined by
$$G^{-1}_{\a\b}(p) =p^2_t\ g_{\a\b} +\bp^2\ k_{\a\b}     \eqno(2.13)$$
where the space components of the momentum are restricted to the
shell (2.9) and the time components are discrete, $p_t=2\pi r/\beta ,
 r\in\bbz$.

To obtain the renormalization group equations it is necessary to
eliminate the ultraviolet cutoff. This can be achieved in the following
way. Firstly, observe that in units of energy the coupling parameters
have the dimension, $D-1$, where $D$ is the number of space dimensions.
It is therefore useful to define a new set of dimensionless couplings
for the effective theory,
$$g_{\a\b} +\Delta g_{\a\b} =\left( {\Lambda\over s}\right)^{D-1}\
  g_{\a\b}(s)                                 \eqno(2.14)$$
and similarly for $k_{\alpha\beta}$. The evolution of these dimensionless
parameters is governed by the differential equation
$${\partial g_{\a\b}(s)\over\partial \ell n\ s} =(D-1)\
  g_{\a\b}(s) +\left( {\Lambda\over s}\right)^{1-D}\
  {\partial\Delta g_{\a\b}\over\partial\ell n\ s}\ .      \eqno(2.15)$$
The cutoff must cancel from the right--hand side when
$\Delta g_{\alpha\beta}$ is expressed as a function of the dimensionless
couplings and the temperature parameter,
$$u ={\Lambda\over s}\ \b\ .                                      \eqno(2.16)$$
It will be verifed in the following that the explicit dependence on
the evolution parameter, $s$, also cancels from the right--hand side of
(2.15).

Although the expressions (2.11) are generally intractable, they simplify
in the case of flag manifolds for which one can obtain fairly explicit
formulae for the renormalization group $\beta$--functions. This is
because the tensors $g_{\alpha\beta}$ and $k_{\alpha\beta}$ are both
diagonal in this case.

The flag manifold associated with a simple group $G$ is defined as the
coset space, $G/H$, where $H$ is the Cartan subgroup of $G$. To study this
case the Cartan--Weyl basis is appropriate. Generators are denoted
$H_j, E_\alpha$ where $\alpha$ is a root. The commutation rules (2.8)
take the form,
$$\eqalignno{
  [H_i,H_j] &=0\cr
  [H_j,E_\a ] &=\alpha_j\ E_\a\cr
  [E_\a ,E_\b ] &=\d_{\a +\b ,0}\ \a^j\ H_j +N_{\a\b}\
  E_{\a +\b}             &(2.17)\cr}$$
where $N_{\alpha\beta}$ is non--vanishing if $\alpha +\beta$ is a root.
The non--vanishing structure constants are therefore,
$$\eqalignno{
  c_{j\a}\ ^\b &=\a_j\ \d_{\a ,\b}\cr
  c_{\a\b}\ ^j &=\a^j\ \d_{\a +\b ,0}\cr
  c_{\a\b}\ ^\g &= N_{\a\b}\ \d_{\a +\b ,\g}\ .    &(2.18)\cr}$$
{}From the hermiticity conditions, $H_j=H^+_j, E_\alpha =E^+_{-\alpha}$
it follows that the structure constants are real and, in particular, that
$N_{\alpha\beta}=N_{-\beta ,-\alpha}$.

Frame components in the tangent space of the flag manifold are now
labelled by the roots. Since the root vectors are all distinct, it follows
from the requirement of $H$--invariance that the tensor $g_{\alpha\beta}$
is diagonal in the sense
$$g_{\a\b} =g_\a\ \d_{\a +\b ,0}                                 \eqno(2.19)$$
and likewise for $k_{\alpha\beta}$. Symmetry implies $g_\alpha =
g_{-\alpha}$ and $k_\alpha =k_{-\alpha}$. Hence, the independent couplings
are equal in number to the (real) dimension of the flag manifold, the
number of roots of $G$.

Our choice of basis vectors, $e_\mu\ ^\alpha$, described in
Appendix A, entails the reality condition
$$(e_\mu\ ^\a )^* =-e_\mu\ ^{-\a}$$
if real coordinates are used. The positivity requirement on $g_{\mu\nu}$
and $k_{\mu\nu}$ therefore implies that the parameters $g_\alpha$
and $k_\alpha$ are {\it negative}
$$g_\a <0,\qquad k_\a <0\ .$$

Now that the propagator (2.13) is diagonal it is straightforward to
evaluate the integrals (2.12). One finds, firstly,
$$\eqalignno{
  G^\alpha &={1\over\b}\ \sum_r\int\left({d\bp\over 2\pi}
  \right)^D\ (p^2_t\ g_\a +\bp^2k_\a )^{-1}\cr
  &=-{K_D\over 2\sqrt{g_\a k_\a}}\ \int^\Lambda_{\Lambda /s}
  dp\ p^{D-2}\ \coth \left({\b p\over 2}\sqrt{{k_\a\over g_\a}}\ \right)
  &(2.20)\cr}$$
where
$$K_D={2(4\pi )^{-D/2}\over\Gamma (D/2)}\ .                 \eqno(2.21)$$
The second integral in (2.12) reduces to
$$\eqalign{
  K^{\g\d\g '\d '}_{\mu\nu} &= (\d_{\g +\g ',0}\ \d_{\d +\d ',0}-
  \d_{\g +\d ',0}\ \d_{\d +\g ',0})\cdot\cr
  &\quad\cdot {1\over\b}\ \sum_r\int\left({dp\over 2\pi}\right)^D\
  p_\mu p_\nu (g_\g\  p^2_t+k_\g\ \bp^2)^{-1}
  (g_\d\ p^2_t +k_\d\ \bp^2)^{-1}\cr}$$
which gives
$$\eqalignno{
  K^{\g\d\g '\d '}_{tt} &=(\d_{\g +\g ',0}\ \d_{\d +\d ',0}-
  \d_{\g +\d ',0}\ \d_{\d +\g ',0})\
  {k_\g\ G^\g -k_\d\ G^\d\over k_\g\ g_\d -k_\d\ g_\g}\cr
  K^{\g\d\g '\d '}_{ij} &= {1\over D}\ \d_{ij}
  (\d_{\g +\g ',0}\ \d_{\d +\d ',0} -
  \d_{\g +\d ',0}\ \d_{\d +\g ',0})\
  {g_\g\ G^\g -g_\d\ G^\d\over g_\g\ k_\d -g_\d\ k_\g}\ .   &(2.22)\cr}$$

There is no need to perform the integral over $p$ in (2.20) since all that
is needed for the renormalization group equations (2.15) is the derivative
$${\partial G^\a\over\partial\ell n\ s} =-\left({\Lambda\over s}
  \right)^{D-1}\ {K_D\over 2\sqrt{g_\a k_\a}}\ \coth
  \left({\beta\Lambda\over 2s}\sqrt{{k_\a\over g_\a}}\ \right)\ .
   \eqno(2.23)$$
The expressions (2.11) for $\Delta g_\alpha$ and $\Delta k_\alpha$ are
linear in $G^\alpha$ and therefore homogeneous of degree zero in the
coupling parameters. The explicit factor $(\Lambda /s)^{D-1}$ in (2.23)
will cancel the corresponding factor in the right--hand side of (2.15).
Furthermore, because of the homogeneity one can replace $g_\alpha$
by $g_\alpha (s)$ and $k_\alpha$ by $k_\alpha (s)$ without
re--introducing any powers of $\Lambda /s$, and without changing the
leading order terms. All cutoff--dependence disappears along with
explicit $s$--dependence and one arrives at the $\beta$--functions.
They depend on $g(s)$, $k(s)$ and the rescaled temperature, i.e.
$$\eqalignno{
  {\partial g_\a\over\partial\ell n\ s} &= (D-1)\ g_\a +
  \a^2g_\a\ K^\a +\sum_\g\ N_{\g\a}
  \Bigl( N_{-\g ,\g +\a}\ g_\a -N_{\g\a}\ g_{\a +\g}\Bigr) K^\g\cr
  &\quad +\sum_\g \left( N_{\a\g}\ g_{\a +\g} +{1\over 2}\
  N_{-\g ,\a +\g}\ g_\a \right)\Bigl( N_{\a\g}\ g_{\a +\g} -
  N_{\a +\g ,-\g}\ g_\a +N_{-\a ,\a +\g}\ g_\g\Bigr)\cdot\cr
  &\qquad\cdot {k_\g\ K^\g -k_{\a +\g}\ K^{\a +\g}\over
  k_\g\ g_{\a +\g} -k_{\a +\g}\ g_\g}\ .
&(2.24)\cr}$$
There is an analogous formula for $\partial k_\alpha /\partial ln\ s$
obtained by exchanging $g$ with $k$ on the right--hand side of (2.24)
and multiplying the last term by ${1\over D}$. In both formula
$K^\alpha$ stands for the function
$$K^\a =-{K_D\over 2\sqrt{g_\a k_\a}}\ \coth\left({u\over 2}
  \sqrt{{k_\a\over g_\a}}\ \right)\ .
\eqno(2.25)$$
The temperature itself evolves according to the trivial formula,
$${\partial u\over\partial\ell n\ s} =-u\ .                     \eqno(2.26)$$

To go further it is necessary to specialize to some particular group.
In the next section we consider the case of $SU(N)$.

\SECTION{THE $SU(N)$ FLAG MANIFOLD}

The generators of $SU(N)$ can be denoted $Q_A\ ^B$, $A,B=1,2.\dots ,N$.
They satisfy the commutation rules
$$[Q_A\ ^B, Q_C\ ^D]=\d^D_A\ Q_C\ ^B-\d^B_C\ Q_A\ ^D     \eqno(3.1)$$
which means that the non--vanishing structure constants equal $\pm\ 1$.
The Cartan generators are identified with the diagonal elements,
$$H_A=Q_A\ ^A\qquad ({\rm no\ sum})                                \eqno(3.2)$$
subject to the constraint $\Sigma\ H_A=0$. The off--diagonal elements are
associated with roots,
$$E_{(A,B)}=Q_A\ B\qquad (A\not= B)\ .
\eqno(3.3)$$
The covariant components of the root vectors, $(A,B)_C$, are then
obtained by comparing the defining formula
$$[H_C,E_{(A,B)}]=(A,B)_C\ E_{(A,B)}$$
with (3.1). One finds
$$(A,B)_C=\d_{BC}-\d_{AC}\ .
\eqno(3.4)$$
In the same fashion one obtains the contravariant components. They
are also given by (3.4). Finally, the sum of two roots is also a root if
the first index on one equals the second index on the other,
$$(A,B)+(B,C)=(A,C)\ .
\eqno(3.5)$$
Hence the non--vanishing components of $N$ are
$$\eqalignno{
  N_{(A,B)(B,C)} &=-1\cr
  &=-N_{(B,C)(A,B)},\qquad (A\not= C)\ .             &(3.6)\cr}$$

Let us now apply this to the formula (2.24) for $\partial g_\a /\partial
\ell n\ s$. It is enough to choose $\a =(1,2)$ in which case the sum over
$\highgamma$ is restricted to roots of the form $(p,1)$ or $(2,p)$,
$p=3,\dots ,N$. One finds,
$$\eqalignno{
  {\partial g_{12}\over\partial\ell n\ s} &=
  (D-1)g_{12}+2g_{12}\ K^{12}+\cr
  &\quad +\sum^N_3\Biggl[ (g_{12}-g_{p2})K^{1p}+(g_{12}-g_{1p})K^{p2}+\cr
  &\quad +(g_{1p}+g_{p2}-g_{12})^2\ {k_{1p}\ K^{1p}-k_{p2}\ K^{p2}
  \over k_{1p}\ g_{p2}-g_{1p}\ k_{p2}}\Biggr]              &(3.7)\cr}$$
and, similarly,
$$\eqalignno{
  {\partial k_{12}\over\partial\ell n\ s} &=
  (D-1)k_{12}+2k_{12}\ K^{12}+\cr
  &\quad +\sum^N_3\Biggl[ (k_{12}-k_{p2})K^{1p}+(k_{12}-k_{1p})K^{p2}+\cr
  &\quad +(k_{1p}+k_{p2}-k_{12})^2\ {g_{1p}\ K^{1p}-g_{p2}\ K^{p2}
  \over g_{1p}\ k_{p2}-k_{1p}\ g_{p2}} {1\over D}\Biggr]        &(3.8)\cr}$$
To obtain the other equations one simply makes the replacements $1\to r,
2\to s$ in (3.7) and (3.8) and takes the sum over $p$ subject to
$p\not= r,s$. For $N=2$ the sums over $p$ are absent from Eqs.(3.7)
and (3.8). It is worthwhile to note that in this case the space
$SU(2)/U(1)=S^2$ admits only a unique (up to scaling) $SU(2)$ invariant
second rank symmetric tensor. Therefore $g_{\mu\nu}=v^2\ k_{\mu\nu}$.
It is easy to verify that in this case if we choose units such that
$v^2=1$ our renormalization group equations reduce to those of
Chakravarty {\it et al.} $^{6)}$.

The Eqs.(3.7) and (3.8) define the evolution of the coupling parameters
in the non--relativistic $SU(N)$ flag manifold $\sigma$--model. In
searching for a fixed point of these equations it is helpful to replace
the parameters, $k_\a$, by velocities,
$$v^2_\a ={k_\a\over g_\a}\ .
\eqno(3.9)$$
The equations then take the form, near $D=1$,
$$\eqalignno{
  {\partial g_{12}\over\partial\ell n\ s} &= (D-1)g_{12}+2K(v_{12})+\cr
  &\quad +\sum_p\Biggl[ {g_{12}-g_{p2}\over g_{1p}}\ K(v_{1p})+
  {g_{12}-g_{1p}\over g_{p2}}\ K(v_{p2})+\cr
  &\quad +{(g_{1p}+g_{p2}-g_{12})^2\over
  g_{1p}\ g_{p2}}\ {v^2_{1p}\ K(v_{1p})-v^2_{p2}\ K(v_{p2})\over
  v^2_{1p} -v^2_{p2}}\Biggr]     &(3.10)\cr
  {\partial v^2_{12}\over\partial\ell n\ s} &= {1\over g_{12}}\
  \sum^N_s\Biggl[ (v^2_{12}-v^2_{1p})\ {g_{1p}\over g_{p2}}\
  K(v_{p2}) +(v^2_{12}-v^2_{p2})\ {g_{p2}\over g_{1p}}\
  K(v_{1p})-\cr
  &\quad - {(v^2_{1p}\ g_{1p}+v^2_{p2}\ g_{p2}-v^2_{12}\ g_{12})^2
  \over g_{1p}\ g_{p2}}\ {K(v_{1p})-K(v_{p2})\over
  v^2_{1p}-V^2_{p2}}\cr
  &\quad - {(g_{1p}+g_{p2}-g_{12})^2\over g_{1p}\ g_{p2}}\
  v^2_{12}\ {v^2_{1p}\ K(v_{1p})-v^2_{p2}\ K(v_{p2})\over
  v^2_{1p}-v^2_{p2}}\Biggr]             &(3.11)\cr}$$
where
$$K(v)={K_D\over 2v}\ \coth {uv\over 2}\ .                     \eqno(3.12)$$

Although these equations are very complicated it is possible to identify
at least one non--trivial fixed point. It appears in the zero temperature
limit, $u\to\infty$, and is given by
$$g_{rs}\to g<0,\qquad v_{rs}\to c>0                           \eqno(3.13)$$
for all pairs $(r,s)$. Firstly, in the limit where all velocities are equal
we have
$$\eqalign{
  {K(v)-K(v')\over v^2-v^{'2}} &\to {1\over 2v}\ {\partial K\over
  \partial v}\cr
  &=-{K_D\over 4v^3}-{K_D\over 2v^3}\ (1+uv)\ e^{-uv} +\dots\cr
  {v^2K(v)-v^{'2}\ K(v')\over v^2-v^{'2}} &\to {1\over 2v}\
  {\partial\over\partial v}\ (v^2K)\cr
  &={K_D\over 4v} +{K_D\over 2v}\ (1-uv)\ e^{-uv} +\dots\cr}$$
where we  have used the low temperature expansion of (3.12),
$$K(v)={K_D\over 2v} +{K_D\over v}\ e^{-uv}+\dots$$
The right--hand side of (3.11) therefore reduces to
$${1\over g_{(1,2)}}\ \sum_p\ {(g_{(1,p)}-g_{(p,2)}-g_{(1,2)})^2\over
  g_{(1,p)}\ g_{(p,2)}}\ K_D\ u\ c^2\ e^{-uc}$$
which vanishes in the limit $u\to\infty$. Next, in the limit where all
the couplings are equal as well, the right--hand side of (3.10) reduces to
$$(D-1)g+2\ {K_D\over 2c} +(N-2)\ {K_D\over 4c}$$
which vanishes for
$$g=-{N+2\over D-1}\ {K_D\over 4c}\ .
\eqno(3.14)$$
The limiting velocity, $c$, is not determined. Its value is a matter
of convention reflecting the choice of units in the original model (1.1).

Finally, to clarify the nature of the fixed point (3.14) we consider the
behaviour of trajectories in its immediate vicinity. To keep the analysis
relatively simply we treat the case, $N=3$ writing
$$g_{12}=g+h_3,\qquad v_{12}=c+u_3
\eqno(3.15)$$
etc., treating $h_\a /g$ and $u_\a /c$ as small quantities. In the linear
approximation, at zero temperature, the evolution equations (3.10)
and (3.11) take the form
$$\eqalignno{
  {\partial\over\partial\ell n\ s}\left({h_1\over g}\right) &=
  {D-1\over 5}\ \left( {h_2+h_3+3h_1\over g} +
  {u_2+u_3+4u_1\over c}\right)\cr
  {\partial\over\partial\ell n\ s}\left({u_1\over c}\right) &=
  {D-1\over 20}\ {u_2+u_3-4u_1\over c}      &(3.16)\cr}$$
and cyclic permutations. In matrix notation,
$${\partial\psi\over\partial\ell n\ s} =M\psi$$
where $\psi$ denotes the column
$$\psi =\left({u_1\over c}, {u_2\over c}, {u_3\over c}, {h_1\over g},
  {h_2\over g}, {h_3\over g}\right)\ .$$
It is straightforward to solve the eigenvalue problem. One finds three
positive (infrared repulsive) eigenvalues and three negative:
$$m_1=m_2={2\over 5}\ (D-1),\ \ m_3=D-1,\ \ m_4=m_5=-
  {1\over 4}\ (D-1),\ \ m_6=-{1\over 10}\ (D-1)\ .$$
The eigenvectors are given by
$$(\psi_1,\dots ,\psi_6) =\left(\matrix{
  0&0&0&-13&-13&-11\cr
  0&0&0&0&26&-11\cr
  0&0&0&13&-13&-11\cr
  1&1&1&12&12&12\cr
  0&-2&1&0&-24&12\cr
  -1&1&1&-12&12&12\cr}\right)\ .$$

Presumably in the case of $SU(N)$ this phenomenon will persist, i.e.
there will be equal numbers of positive and negative eigenvalues.

\SECTION{DISCRETE SYMMETRIES}

At the fixed point discussed in the previous section, the coupling
parameters degenerate in that $g_\a$ and $k_\a$ are independent of
$\a$. One therefore expects to find some enhancement of the symmetries
of the system. It is unlikely that new continuous symmetries would
emerge since the target space is homogeneous and the action of the
group $G$ on this space is already, in a certain sense, maximal. It is
determined by the dimensionality and topology of the space. Discrete
symmetries, on the other hand, cannot be excluded. Indeed, it is quite
easy to show, at least for the $SU(N)$ flag manifold, that the critical
Lagrangian admits a group of permutations acting on the target space.
To demonstrate this we begin with a brief discussion of the
automorphisms of $G$.

We are interested in transformations,
$$Q_{\ha}\to S_{\ha}\ ^{\hb}\ Q_{\hb}                             \eqno(4.1)$$
that leave invariant the algebra of $G$, i.e.
$$S_{\ha}\ ^{\ha_1}\ S_{\hb}\ ^{\hb_1}\ c_{\ha_1\hb_1}\ ^{\hg} =
  c_{\ha\hb}\ ^{\hg_1}\ S_{\hg_1}\ ^{\hg}$$
or, equivalently,
$$S^{-1}q_{\ha}\ S=S_{\ha}\ ^{\hb}\ q_{\hb}                       \eqno(4.2)$$
where $q_{\ha}$ denotes the adjoint representation,
$$(q_{\ha})_{\hb}\ ^{\hg} =c_{\hb\ha}\ ^{\hg}\ .               \eqno(4.3)$$
The coset space, $G/H$, is parametrized by a set of fields $\phi^\a$.
For example, one might choose the exponential parametrization,
$$L_\phi =e^{\phi^\a Q_\a}\ .$$
In the adjoint representation this reads
$$D_{\ha}\ ^{\hb} (L_\phi )=(e^{\phi^{\g} q_{\g}})_{\ha}\ ^{\hb}\ .$$
Using this representation it is clear that the automorphism, $S$, can be
made to act on the coordinates, $\phi$, if it leaves invariant the algebra
of $H$, i.e. if $S_{\ha}\ ^{\hb}$ is block diagonal,
$$\eqalignno{
  S^{-1}\ q_{\ba}\ S &=S_{\ba}\ ^{\bb}\ q_{\bb}\cr
  S^{-1}\ q_\a\ S &=S_\a\ ^\b\ q_\b\ .         &(4.4)\cr}$$
Writing
$$S^{-1}\ D(L_\phi )S=D(L_{\phi '})                        \eqno(4.5)$$
one defines the linear transformation
$$\phi^\a\to\phi^{'\a}=\phi^\b\ S_\b\ ^\a\ .                \eqno(4.6)$$
With the 1--forms $e^{\ha}$ defined by
$$L^{-1}_\phi\ dL_\phi =d\phi^\mu\ e_\mu\ ^{\ha}(\phi )\ Q_{\ha}$$
it follows from (4.5) and (4.6) that
$$e_\mu\ ^{\ha}(\phi ')=(S^{-1})_\mu\ ^\nu\ e_\nu\ ^{\hb}(\phi )\
  S_{\hb}\ ^{\ha}\ .                                        \eqno(4.7)$$
This implies that the Lagrangian (1.1) is invariant under those
automorphisms that leave invariant the frame components of the metric
tensors,
$$\eqalignno{
  S_\a\ ^\g\ S_\b\ ^\d\ g_{\g\d} &= g_{\a\b}\cr
  S_\a\ ^\g\ S_\b\ ^\d\ k_{\g\d} &= k_{\a\b}\ .          &(4.8)\cr}$$
If these conditions are satisfied it is straightforward to show that the
Noether currents transform according to (4.1) as one would expect.

In the case of $SU(N)$ with generators, $Q_A\ ^B$, as described in
Sec.3 it can be verified that the automorphisms include the $N!$
permutations of the indices,
$$Q_A\ ^B\to Q_{\pi_A}\ ^{\pi_B}\ .                          \eqno(4.9)$$
These permutations are also automorphisms of the Cartan algebra. Since
this group acts by permuting the root vectors, it follows that the
Eqs.(4.8) will be satisfied when the coupling parameters, $g_\a$ and
$k_\a$, are independent of $\a$. Hence, the permutations emerge as a
good symmetry at the fixed point. However, it should be noted that this
group action has fixed points.

\SECTION{DISCUSSION}

The model whose infrared behaviour is studied in this paper is itself
obtained as a classical long wavelength limit of a generalized spin
system on a lattice. We have simply quantized this model and extracted the
resulting 1--loop contributions to the $\beta$--functions in the usual
way. One may object that the quantum nonlinear $\sigma$--model
arrived at in this way has nothing to do with the original spin system
since the intermediate step involves the neglect of quantum phenomena
which may be important. To this objection we can only respond by
arguing that the neglected quantities -- associated with factor ordering
ambiguities -- are short distance effects and should therefore be
irrelevant for the infrared behaviour. Such ordering effects are in any
case discarded in computations of quantum corrections to $\sigma$--models.
Indeed, the opinion seems to be widely shared that the long wavelength
properties of the quantized $\sigma$--model do in fact represent those
of the spin system.

Another objection could be that our use of the momentum shell technique
for computing $\beta$--functions is not compatible with the
requirements of group invariance. Feynmann integrals made finite by a
simple momentum space cutoff generally fail to satisfy the appropriate
Ward identities. The amplitudes associated with inherently divergent
graphs such as those of Fig.1 will turn out to be tensors with respect to
transformations acting on the background fields only if they are
regularized covariantly. This we have not done. A careful evaluation
using our cutoff prescriptions would reveal some unwanted dependence
on the background spin connection. These terms we have discarded,
(they actually disappear in the case of one space dimension at zero
temperature, otherwise not) believing them to be artifacts. However,
the approach clearly leaves something to be desired. Nevertheless, to
the order we have considered we believe that these problems do not
affect our results. In support of this in Appendix B, we have argued
that the manifestly covariant dimensional regularization will
produce the same zero temperature $\b$--function when $D=1+\v$.

The cutoff problems associated with $\sigma$--models are of course
pseudo--problems from the viewpoint of the original spin system. As
we pointed out in Sec.1, momentum space is compact for the spin
system. A fully consistent approach to obtaining the long wavelength
properties would be to proceed by ``decimation'', i.e. integrating out
the variables associated with alternate sites, for example. A formulation
of this kind is under development at present.
\vfill\eject

\noindent{\bf APPENDIX A\qquad The normal coordinate expansion}

To obtain the expansion (2.4) it is necessary to establish some simple
properties of coset manifolds, $G/H$, and choose an appropriate
connection form. More particularly, one needs to specify the general
structure of $G$--invariant tensors and to find the connection with
respect to which they are covariantly constant.  The curvature and torsion
tensors associated with this connection can then be evaluated.

The manifold can be coordinatized by choosing a representative element,
$L_\phi\in G$, from each left coset of $G$. The action of $G$ on the
manifold is then represented in the form $\phi\to\phi '$ where
$$g\ L_\phi =L_{\phi '} h                                   \eqno(A.1)$$
with $g\in G$ and $h\in H$. Consider the 1--form
$$L^{-1}_\phi\ dL_\phi =e^\a\ Q_\a +e^{\ba}\ Q_{\ba}       \eqno(A.2)$$
where $Q_{\ba}$ spans the algebra of $H$ and $Q_\a$ the remaining
part. The 1--forms $e^\a =d\phi^\mu\ e_\mu\ ^\a (\phi )$ define a
frame basis for the tangent space of $G/H$. They transform covariantly
under the action of $G$,
$$e^\a (\phi ')=e^\b (\phi )\ D_\b\ ^\a (h)$$
where $h=h(\phi ,g)$ is determined by (A.1). The representation $h\to
D(h)$ is determined by the adjoint representation of $G$, restricted to
the subgroup $H$. It follows that tensors like $g_{\mu\nu}(\phi )$ are
$G$--invariant if their frame components, $g_{\a\b}(\phi )$, are
$H$--invariant constants.

The Maurer--Cartan equations follow from (A.2) together with the
commutation rules (2.8),
$$\eqalignno{
  de^\a +e^\b\wedge e^{\bg}\ c_{\b\bg}\ ^\a  &={1\over 2}\
  e^\g\wedge e^\b\ c_{\b\g}\ ^\a      &(A.3)\cr
  de^{\ba} -{1\over 2}\ e^{\bg}\wedge e^{\bb}\ c_{\bb\bg}\ ^{\ba} &=
  {1\over 2}\ e^\g\wedge e^\b\ c_{\b\g}\ ^{\ba}  \ .    &(A.4)\cr}$$
The first of these equations serves to identify a connection,
$$\omega_\b\ ^\a =-e^{\bg}\ c_{\b\bg}\ ^\a                  \eqno(A.5)$$
and its associated torsion,
$$T_{\b\g}\ ^\a =-c_{\b\g}\ ^\a\ .
\eqno(A.6)$$

As usual, one can define the components of the connection in the
coordinate basis, $\Gamma_{\mu\nu}\ ^\lambda$, such that the frames
are covariantly constant,
$$\eqalignno{
  0 &=\nabla_\mu\ e_\nu\ ^\a\cr
  &=\partial_\mu\ e_\nu\ ^\a -\Gamma_{\mu\nu}\ ^\lambda\
  e_\lambda\ ^\a +e_\nu\ ^\b\ \omega_{\mu\b}\ ^\a\ .           &(A.7)\cr}$$
The curvature tensor is defined in the coordinate basis by
$$\eqalignno{
  R_{\mu\nu\l}\ ^\rho &= \partial_\mu\ \Gamma_{\nu\l}\ ^\rho -
  \partial_\nu\ \Gamma_{\mu\l}\ ^\rho -\Gamma_{\mu\l}\ ^\s\
  \Gamma_{\nu\s}\ ^\rho +\Gamma_{\nu\l}\ ^\s\
  \Gamma_{\mu\s}\ ^\rho\cr
  &=e_\l\ ^\a\left(\partial_\mu\ \omega_\nu -\partial_\nu\
  \omega_\mu -[\omega_\mu ,\omega_\nu ]\right)_\a\ ^\b\
  e_\b\ ^\rho\cr
  &=e_\l\ ^\a\ e_\mu\ ^\g\ e_\nu\ ^\d\ R_{\g\d\a}\ ^\b\
  e_\b\ ^\rho                          &(A.8)\cr}$$
where, from the Maurer--Cartan equation (A.4), the frame components are
$$R_{\g\d\a}\ ^\b =c_{\g\d}\ ^{\bs}\ c_{\a\bs}\ ^\b\ .      \eqno(A.9)$$
The torsion tensor is obtained in the coordinate basis by using (A.7)
and (A.3),
$$\eqalignno{
  T_{\mu\nu}\ ^\a &=\Gamma_{\mu\nu}\ ^\l -\Gamma_{\nu\mu}\ ^\l\cr
  &=\left(\partial_\mu\ e_\nu\ ^\a +e_\nu\ ^\b\ \omega_{\mu\b}\ ^\a -
  (\mu\leftrightarrow \nu )\right) e_\a\ ^\l\cr
  &=-e_\mu\ ^\b\ e_\nu\ ^\g\ c_{\b\g}\ ^\a\ e_\a\ ^\l\ .  &(A.10)\cr}$$

The covariant constancy of a tensor such as
$$k_{\mu\nu}(\phi )=e_\mu\ ^\a (\phi )\ e_\nu\ ^\b (\phi )\
  k_{\a\b} (\phi )$$
is expressed in the equations
$$\eqalign{
  0 &=\partial_\l\ k_{\mu\nu} -\Gamma_{\l\mu}\ ^\rho\ k_{\rho\nu} -
  \Gamma_{\l\nu}\ ^\rho\ k_{\mu\rho}\cr
  &=e_\mu\ ^\a\ e_\nu\ ^\b \left(\partial_\l\ k_{\a\b} -
  \omega_{\l\a}\ ^\g\ k_{\g\b} -\omega_{\l\b}\ ^\g\
  k_{\a\g}\right)\cr
  &=e_\mu\ ^\a\ e_\nu\ ^\b\left(\partial_\l\ k_{\a\b} +
  e_\l\ ^{\bd}\ c_{\a\bd}\ ^\g\ k_{\g\b} +e_\l\ ^{\bd}\
  c_{\b\bd}\ ^\g\ k_{\a\g}\right)\cr}$$
or, in effect,
$$\eqalignno{
  \partial_\l\ k_{\a\b} &= 0\cr
  c_{\a\bd}\ ^\g\ k_{\g\b} +c_{\b\bd}\ ^\g\ k_{\a\g} &= 0   &(A.11)\cr}$$
which means that the frame components are $H$--invariant and
independent of $\phi$. This generalizes to tensors of any rank. Notice
that the curvature and torsion tensors, in particular, are covariantly
constant.

Turning now to the problem of developing a covariant expansion around an
arbitrary background we use a method due to
Alvarez--Gaum\'e {\it et al.} $^{7)}$. As explained
in Sec.2 the idea is to introduce a field $\highgamma^\mu (x,u)$ that
interpolates between an arbitrary configuration, $\phi^\mu (x)=
\highgamma^\mu (x,1)$, and some fixed background configuration,
$\bar\phi^\mu (x)=\highgamma^\mu (x,0)$. In principle, one should be
able to express $\phi^\mu (x)$ as a Taylor expansion of $\highgamma^\mu$
and its derivatives with respect to $u$, evaluated at $u=0$. If
$\highgamma^\mu (x,u)$ is made to satisfy a second order differential
equation in $u$, then the higher derivatives can all be expressed in
terms of the first derivative at $u=0$, which becomes the new
independent variable. The aim is to maintain covariance with respect
to coordinate transformations on the target manifold.

The simplest covariant second order differential equation is the geodesic
equation,
$$\partial^2_u\ \highgamma^\mu +\partial_u\highgamma^\l\
  \partial_u\highgamma^\nu\ \Gamma_{\nu\l}\ ^\mu (\highgamma )=0$$
where $\Gamma_{\nu\lambda}\ ^\mu$ is a connection. It is useful to define
the tangent vector
$$\xi^\mu =\partial_u\highgamma^\mu                     \eqno(A.12)$$
and write the geodesic equation in the form
$$\nabla_u\ \xi^\mu =0\ .                                      \eqno(A.13)$$
Partial derivatives with respect to spacetime coordinates, $\partial_a
\highgamma^\mu$ are also covariant. They can be used to define a covariant
derivative of the tangent vector
$$\nabla_a\ \xi^\mu =\partial_a\ \xi^\mu +\partial_a\
  \highgamma^\l\ \xi^\nu\ \Gamma_{\l\nu}\ ^\mu\ .          \eqno(A.14)$$
Higher derivatives follow in an obvious way,
$$\eqalignno{
  \nabla_u \partial_a\highgamma^\mu &=\partial_u\partial_a
  \highgamma^\mu +\xi^\l\partial_a\highgamma^\nu
  \Gamma_{\l\nu}\ ^\mu =\cr
  &=\partial_a\xi^\mu +\xi^\l\partial_a\highgamma^\nu
  \Gamma_{\l\nu}\ ^\mu\cr
  &=\nabla_a\xi^\mu +\xi^\l\partial_a\highgamma^\nu
  T_{\l\nu}\ ^\mu\ ,    &(A.15)\cr
  \nabla_u\nabla_a\xi^\mu &= \partial_u(\nabla_a\xi^\mu )+\xi^\l\nabla_a
  \xi^\nu\Gamma_{\l\nu}\ ^\mu\cr
  &= \dots\cr
  &= -\partial_a\highgamma^\rho\ R_{\rho\l\nu}\ ^\mu\ \xi^\nu\xi^\l\ .
  &(A.16)\cr}$$
The curvature and torsion tensors that arise here are defined by
(A.8)--(A.10). They are covariantly constant,
$$\nabla_u\ R_{\rho\l\nu}\ ^\mu =0=\nabla_u\ T_{\l\nu}\ ^\mu\ ,
                                             \eqno(A.17)$$
a feature which simplifies the higher derivatives of $\partial_a
\highgamma^\mu$ and $\nabla_a\xi^\mu$.

The covariant derivative is of course distributive and, acting on an
invariant it reduces to the ordinary derivative. For example, on the
scalar product of two tensors,
$$\partial_u(A\cdot B)=\nabla_uA\cdot B+A\cdot\nabla_uB\ .$$
The Lagrangian term, $(1/2)k_{\mu\nu}(\highgamma )\partial_j
\highgamma^\mu\partial_j\highgamma^\nu$, is a scalar with respect to
transformations in the target space and its first few derivatives are
easily calculated.
\vfill\eject
$$\eqalignno{
  \partial_u\left({1\over 2}\ k_{\mu\nu}\ \partial_j\highgamma^\mu\
  \partial_j\highgamma^\nu \right) &= k_{\mu\nu}\ \partial_j
  \highgamma^\mu \left(\nabla_j\xi^\nu +\xi^\l\ \partial_j
  \highgamma^\rho\ T_{\l\rho}\ ^\nu\right)\cr
  \partial^2_u\left({1\over 2}\ k_{\mu\nu}\ \partial_j
  \highgamma^\mu\partial_j\highgamma^\nu\right) &=
  k_{\mu\nu}\left(\nabla_j\xi^\mu +\xi^\kappa\partial_j
  \highgamma^\s T_{\kappa\s}\ ^\mu\right)\left(\nabla_j\xi^\nu +
  \xi^\l\partial_j\highgamma^\rho T_{\l\rho}\ ^\nu\right) +\cr
  &\quad + k_{\mu\nu}\partial_j\highgamma^\mu\left[ -
  \partial_j\highgamma^\rho\ R_{\rho\l\s}\ ^\nu \xi^\l\xi^\s +
  \xi^\l\left(\nabla_j\xi^\rho +\xi^\s\partial_j\highgamma^\tau
  T_{\s\tau}\ ^\rho\right) T_{\l\rho}\ ^\nu\right]\cr
  &= k_{\mu\nu}\ \nabla_j\xi^\mu\ \nabla_j\xi^\nu\cr
  &\quad + 2\left( T_{\nu\rho}\ ^\l\ k_{\l\mu} +
  {1\over 2}\ T_{\nu\mu}\ ^\l\ k_{\l\rho}\right)\xi^\nu\nabla_j\xi^\mu\
  \partial_j\highgamma^\rho\cr
  &\quad +\left( -R_{\rho\mu\nu}\ ^\tau \ K_{\tau\s} +
  k_{\rho\tau}\ T_{\nu\s}\ ^\l\ T_{\mu\l}\ ^\tau +
  k_{\l\tau}\ T_{\nu\rho}\ ^\l\ T_{\mu\s}\ ^\tau\right)
  \xi^\mu\xi^\nu\partial_j\highgamma^\rho\partial_j
  \highgamma^\s\cr
  &                      &(A.18)\cr}$$
where we have used (A.15) and (A.16) together with the covariant
constancy of $k_{\mu\nu}$.

Analogous formulae for derivatives of the other Lagrangian term,\hfil\break
$(1/2)g_{\mu\nu} (\highgamma ) \partial_t \highgamma^\mu \partial_t
\highgamma^\nu$, are obtained from these by replacing $k$ with $g$
and $\partial_j$ with $\partial_t$. It remains only to substitute these
expressions, evaluated at $u=0$, i.e. with $\highgamma^\mu$ replaced
by $\bar\phi^\mu$, in the Taylor expansion. On setting $u=1$ the
expansion (2.4) is obtained.

Finally, concerning the choice of basis vectors for the tangent space of
the flag manifold, we write (A.2) in the form
$$L^{-1}_\phi\ dL_\phi =\sum_{roots}\ e^\a\ E_\a +e^j\ H_j \eqno(A.19)$$
where $E^+_\a =E_{-\a}$ and $H^+_j=H_j$. Since $L^{-1}dL$ is
antihermitian this implies the reality conditions
$$e^{\a *} =-e^{-\a}\quad {\rm and}\quad e^{j*} =-e^j\ .$$
If the coordinates $\phi^\mu$ are real, as we have assumed throughout
this paper, then
$$\eqalignno{
  e_\mu\ ^\a (\phi )^* &=-e_\mu\ ^{-\a}(\phi )\cr
  e_\mu\ ^j(\phi )^* &=-e_\mu\ ^j(\phi )\ .       &(A.20)\cr}$$
The positivity of $g_{\mu\nu}(\phi )$ therefore takes the form
$$g_{\mu\nu}\ d\phi^\mu\ d\phi^\nu =-\Sigma\ g_\a\vert d\phi^\mu\
  e_\mu\ ^\a\vert^2 >0                                  \eqno(A.21)$$
implying that the coefficients $g_\a$ are negative. Also, the spin
connection is pure imaginary,
$$\eqalignno{
  \omega_{\mu\a}\ ^\b &=-e_\mu\ ^j\ c_{\a j}\ ^\b\cr
  &=e_\mu\ ^j\ \a_i\ \d_{\a ,\b}\ .             &(A.22)\cr}$$
\vfill\eject

\noindent{\bf APPENDIX B\qquad The renormalization group at $T=0$}

In this appendix we would like to discuss some technical matters
concerning the various ways of regularizing the divergent integrals.

The 1--loop expressions for $\Delta g_\a$ and $\Delta k_\a$ are given by
$$\eqalignno{
  \Delta g_\a &=\alpha^2g_\a\ G^\a +\sum_\g\ N_{\g\a}\left(
  N_{-\g ,\g +\a}\ g_\a-N_{\g ,\a}\ g_{\a +\g}\right)G^\g\cr
  &\quad +\sum_\g\left( N_{\a\g}\ g_{\a +\g}+{1\over 2}\
  N_{-\g ,\a +\g}\ g_\a\right)\Bigl( N_{\a\g}\ g_{\a +\g} -
  N_{\a +\g ,-\g}\ g_\a +\cr
  &\quad + N_{-a ,\a +\g}\ g_\g\Bigr) \ {k_\g\ G^\g -k_{\a +\g}
 \  G^{\a +\g}\over k_\g\ g_{\a +\g} -k_{\a +\g}\ g_\g}           &(B.1)\cr}$$
$$\eqalignno{
  \Delta k_\a &=\alpha^2k_\a\ G^\a +\sum_\g\ N_{\g\a}\left(
  N_{-\g ,\g +\a}\ k_\a-N_{\g ,\a}\ k_{\a +\g}\right)G^\g\cr
  &\quad +{1\over D}\sum_\g\left( N_{\a\g}\ k_{\a +\g}+{1\over 2}\
  N_{-\g ,\a +\g}\ k_\a\right)\Bigl( N_{\a\g}\ k_{\a +\g} -
  N_{\a +\g ,-\g}\ k_\a +\cr
  &\quad + N_{-\a ,\a +\g}\ k_\g\Bigr) \ {g_\g\ G^\g -g_{\a +\g}\
  G^{\a +\g}\over g_\g\ k_{\a +\g} -g_{\a +\g}\ k_\g}\ .       &(B.2)\cr}$$
It is indeed through the differentiation of these expressions with respect
to $\ell n\ s$ that we obtained equations such as (2.24). At zero $T$ the
propagator $G^\alpha$ is given by
$$G^\a ={1\over g_\a}\int \left({dp\over 2\pi}\right)^{D+1}\
  {1\over p^2_0+v^2_\alpha\ \bp^2}                               \eqno(B.3)$$
where $v^2_\a ={k_\a\over g_\a}$. The scheme which we have adopted
in Sec.2 defines this propagator by integrating $p_0$ from $-\infty$
to $+\infty$
while restricting the $\bp$--integrals to the shell ${\Lambda\over s}
\leq\vert\ \bp\vert\leq\Lambda ;s>1$. This scheme clearly breaks the
$D+1$ dimensional rotational symmetry that the classical action
would have whenever $v_\a$ is independent of $\a$. This scheme is
suitable in the quantum domain only for $T>0$ $(T\not= 0)$ where
$p_0$ becomes a discrete variable and the $(D+1)$--dimensional
symmetry is broken to a $D$--dimensional one. On the other hand at
$T=0$ it is possible alternatively to integrate over a $(D+1)$--dimensional
shell
$${\Lambda\over s}\leq p^2_0+\vert\bp\vert^2\leq\Lambda\ .$$
In this case it is not hard to see that if $v_\a$ is independent of $\a$,
the $(D+1)$--dimensional rotational symmetry will be left intact.

A serious defect of the momentum shell technique is its failure to
respect the Ward identities associated with the $G$--invariance. This
is due to the fact that a simple cutoff, $\Lambda$, is introduced in the
integration over $p$. For $D=1$ and to the order that we are considering
it is probably a happy accident that the explicit connection--dependent
terms cancel out. However, in order to be confident about the
correctness of our $\b$--functions we have verified that the same
functions can be otbained using dimensional regularization, which
respects the symmetries of the problem. To this end it is necessary
only to identify the singular part of $G^\a$ as $D\to 1$. This is given by
$$G^\a =-{1\over 2\pi}\ {1\over\sqrt{g_\a k_\a}}\ {1\over D-1} +
  {\rm regular\ terms}\ .$$
Using this expression for $G^\a$ in Eqs.(B.1) and (B.2), it is easy to verify
that the minimal subtraction prescription will reproduce the same
$\b$--function as the $T\to 0$ limit in Eqs.(3.7) and (3.8).
\vfill\eject

\centerline{REFERENCES}
\bigskip

\item{1)}
S. Randjbar--Daemi, Abdus Salam and J. Strathdee,
``Generalized spin systems and
$\sigma$--models'', ICTP, Trieste, preprint IC/92/294 (1992).

\item{2)}
See for example,\hfil\break
I. Affleck, in {\it Strings, Fields and Critical Phenomena}, Les Houches
Summer School 1988, Session XLIX, eds. E. Brezin and J. Zinn--Justin,
North Holland (1990);\hfil\break
E. Fradkin, {\it Field Theories of Condensed Matter Systems},
Addison--Wesley, Redwood City (1991);\hfil\break
E. Manousakis, Rev. Mod. Phys. {\bf 63}, 1 (1991).

\item{3)}
F.D.M. Haldane, J. Appl. Phys. {\bf 57}, 3359 (1985).

\item{4)}
J. Zinn--Justin, {\it Quantum Field Theory and Critical Phenomena},
Clarendon Press, Oxford (1989).

\item{5)}
C. Hull, in {\it Super Field Theories}, eds. H.C. Lee et al., Plenum, New
York (1987).

\item{6)}
S. Chakravarty, B.I. Halperin and D.R. Nelson, Phys. Rev. {\bf B39}, 2344
(1989).

\item{7)}L. Alvarez--Gaum\'e, D.Z. Freedman and S. Mukhi, Ann. Phys.
{\bf 134}, 85 (1981);\hfil\break
See also Ref.5 for a review.

\item{8)}
J. Kogut and K. Wilson, Phys. Rep. {\bf 12C}, 76 (1973);\hfil\break
S.K. Ma, {\it Modern Theory of Critical Phenomena}, Benjamin,
Philadelphia (1986).
\vfill\eject
\end